# Landau level spectroscopy of relativistic fermions with low Fermi velocity in $Bi_2Te_3$ three-dimensional topological insulator

A. Wolos

*Institute of Physics, Polish Academy of Sciences, Al. Lotnikow 32/46, 02-668 Warsaw, Poland.*
*Faculty of Physics, University of Warsaw, ul. Hoza 69, 00-681 Warsaw, Poland.*

S. Szyszko, A. Drabinska, and M. Kaminska

*Faculty of Physics, University of Warsaw, ul. Hoza 69, 00-681 Warsaw, Poland.*

S. G. Strzelecka, A. Hruban, A. Materna, and M. Piersa

*Institute of Electronic Materials Technology, ul. Wolczynska 133, 01-919 Warsaw, Poland.*

ABSTRACT:

X-band microwave spectroscopy is applied to study the cyclotron resonance in $Bi_2Te_3$ exposed to ambient conditions. With its help, intraband transitions between Landau levels of relativistic fermions are observed. The Fermi velocity equals to 3260 m/s, which is much lower than has been reported in the literature for samples cleaved in vacuum. Simultaneous observation of bulk Shubnikov - de Haas oscillations by contactless microwave spectroscopy allows determination of the Fermi level position. Occupation of topological surface states depends not only on bulk Fermi level but also on the surface band bending.

In the last five years a new class of topological quantum states has emerged, referred to as topological insulators.[1,2,3,4,5,6,7,8,9,10] Topological properties of the family of bismuth compound crystals, i.e. $Bi_2Te_3$,[5] $Bi_2Se_3$,[8] and $Bi_2Te_2Se$,[9] have been confirmed in spectacular angle resolved photoemission spectroscopy (ARPES) and scanning tunneling microscopy and spectroscopy[11] experiments. Other experiments, including electric transport or optical measurements, are difficult to perform because of high bulk conductance due to native crystal lattice defects.[6,8,12] The defects responsible for a high bulk carrier concentration are most probably Bi antisites in $Bi_2Te_3$ leading to typically highly metallic p-type, and Se vacancies in $Bi_2Se_3$ giving n-type conductivity.[6,13,14] In order to suppress the contribution of the conducting bulk states and expose the surface states in the experiment, very thin samples with the thickness of a few quintet layers are typically studied.[15,16] An alternative approach is to use surface sensitive experimental techniques or techniques disregarding the metallic bulk. One of them is the already mentioned ARPES, which probes surface regions from a few up to tens of nm. In this communication we describe the results of microwave spectroscopy measurements performed in an X-band (9.5 GHz) resonator. The microwave penetration depth in the applied experimental conditions ranges from 4 μm to 40 μm. The technique can actually probe independently both bulk and surface properties, being partly insensitive to the detrimental influence of the metallic bulk conductivity. Similar approach has been recently applied to study surface states in $Bi_2Se_3$ topological insulator, where 2D cyclotron resonance has been revealed thanks to the application of the microwave cavity transmission technique.[17]

The $Bi_2Te_3$ crystal was grown by the vertical Bridgman method, with an excess of tellurium in the melt in order to suppress the formation of Bi antisite defects. Growth details have been described in Ref. 18. Segregation effects, characteristic of the Bridgman method, caused the gradient of stoichiometry along the crystal growth direction. For a given temperature, the equilibrium composition of the liquid phase is typically different than of the solid phase. In the case of $Bi_2Te_3$, this causes enrichment of the liquid phase in tellurium during the growth process, and formation of the concentration gradient of defects related to tellurium excess. The crystal, highly p-type at the crystal seed due to large concentration of Bi antisites (p = $1\times10^{20}$ cm$^{-3}$ at room temperature), changed its conductivity to n-type at its end (n = $3\times10^{19}$ cm$^{-3}$ at room temperature). The crystal was cut with a wire saw and cleaved with the razor edge to obtain samples with surface area not larger than 4 mm by 4 mm, to fit to the resonance cavity. Mirror-like surfaces perpendicular to the $Bi_2Te_3$ **c**-axis were obtained. The thickness of all the samples was typically about 0.5 mm. Three samples from p-type region of the crystal were prepared for the studies, cut at 73 mm (A), 80 mm (B), and 107 mm (C) from the seed. The samples were kept at ambient conditions.

The microwave spectroscopy experiment was performed using a Bruker ELEXSYS E580 electron spin resonance (ESR) spectrometer operating in X-band (9.5 GHz), with a $TE_{102}$ resonance cavity. The temperature was lowered down to 2.5 K using an Oxford continuous-flow cryostat. The magnetic field was swept up to 1.7 T. Due to the use of the magnetic field modulation and the lock-in detection technique, the recorded spectra take the form of the first derivative of the microwave power absorption.

Two kinds of signals were observed, the Shubnikov - de Haas (SdH) oscillations and the cyclotron resonance. The SdH oscillations were visible in the high magnetic field range for samples A, B, and C, see Fig. 1 and Table 1. The SdH signal can be detected with the ESR spectrometer due to non-resonant changes in sample conductivity causing variation of the cavity quality factor. The contactless method of the SdH oscillations detection has been earlier applied for a high-mobility two-dimensional electron gas present at the $GaN/Al_xGa_{1-x}N$ interface[19] or in $In_xGa_{1-x}As$ quantum wells.[20]

The SdH signal shows characteristic periodicity in 1/B (B is the external magnetic field). The Fourier transform of the signal (Fig. 1 (b)) reveals only one frequency of the oscillations, equal to 14.2 T, 14.0 T, and 7.0 T for samples A, B, and C respectively (for **B** || **s**). Table 1. summarizes the parameters. The decrease of the SdH frequency, when the sample is taken away from the seed, is consistent with increasing the excess of tellurium in the melt during the growth process what causes systematic suppression of acceptor-like Bi antisite defect formation. The SdH oscillations in $Bi_2Te_3$ are well visible for the magnetic field oriented parallel to the $Bi_2Te_3$ bisectrix **s**-axis (in-plane of the cleavage surfaces), and up to about 30 degrees from this orientation towards the $Bi_2Te_3$ **c**-axis. The frequency of the oscillations increases slightly while rotating the sample, the onset moves towards higher magnetic fields, Fig 1(a) and (c). The angular dependence of the oscillations excludes the possibility of their originating from the 2D surface states, the oscillations should vanish for **B** in-plane of the cleavage surface in that case, and allows assigning them to the bulk holes.

The structure of the Fermi surfaces in p-type $Bi_2Te_3$ (bulk) has been studied by first principles calculations, revealing a rather complicated picture of irregular elongated surfaces, located in the mirror plane of the $Bi_2Te_3$ Brillouin zone, tilted by 21 degrees with respect to the crystal principal axes.[21] At least 6 hole pockets exist in the Brillouin zone. A simplified 6-valley model of ellipsoidal Fermi surfaces with non-parabolic dispersion has been applied by Köhler in 1976 to explain the pattern of the SdH oscillations in $Bi_2Te_3$.[22,23] Significant non-parabolicity effects have been visible for the Fermi energy above 20 meV, while at around 25 meV the slope of $m_c(E_F)$ has been infinite ($m_c$ is the cyclotron mass, $E_F$ – the Fermi energy counted from the bulk valence band edge, it increases when increasing hole concentration). The band edge cyclotron masses have been deduced by Köhler as equal to 0.080 $m_0$ for **B** || **c**, and 0.058 $m_0$ for **B** || **s**. For the Fermi energy $E_F$ = 21.6 meV, the cyclotron masses increase up to 0.102 $m_0$ for **B** || **c** and 0.075 $m_0$ for **B** || **s**.

In general, the SdH pattern in $Bi_2Te_3$ is very complex, consisting of up to three periods at the arbitrary orientation of the magnetic field, with different amplitudes and field and temperature dependencies.[22] In our experiment, due to the magnetic field limited to 1.7 T, we can only observe components characterized by the highest mobility and thus by the lowest cyclotron mass. This happens for orientations close to **B** || **s**. Frequencies of the SdH oscillations, F = 14.2 T and 14.0 T, observed for samples A and B respectively, are very close to the value observed by Köhler when $m_c$ equals to 0.075 $m_0$. A lower value of the SdH frequency in sample C, F = 7.0 T, suggests that the corresponding cyclotron mass is of the band edge (0.058 $m_0$ for **B** || **s,** after Köhler).

The SdH period $\Delta(1/B)$, the cyclotron mass $m_c$, and Fermi energy $E_F$, are related through the equation:[22]

$$\Delta(1/B) = e\hbar/(m_c E_F), \qquad \text{Eq.1.}$$

where e is the electron charge and $\hbar$ is a Planck constant over $2\pi$. Using this relation, we can roughly determine the position of the Fermi level in our samples, assuming the cyclotron mass of 0.075 $m_0$ for A and B samples and the band edge cyclotron mass 0.058 $m_0$ for sample C. We obtain $E_F$ equal to about 22 meV for samples A and B respectively, and $E_F$ = 14 meV for sample C. Table 1 collects the estimated parameters.

In sample C, a strong signal shown in Fig.2(a) was detected next to the SdH oscillations. A large amplitude of the signal and a strong dependence of the amplitude on the position in the resonance cavity (optimized for measurements of magnetic dipole transitions) suggests the microwave electric field-driven cyclotron resonance. Cyclotron resonance-related lines are commonly detected in ESR cavities both in bulk materials[24] as well as in 2D systems like Si/SiGe quantum wells[25] or GaN/AlGaN heterostructures.[19] We should mention here that the signal did not show any signs of aging within a time scale of a month.

The spectrum in Fig. 2(a). shows clear structures for **B || c,** and becomes flat for **B || s** (in-plane of the cleavage surfaces). Three resonance lines can be distinguished, marked as L1, L2, and L3, at the magnetic field equal to 0.61 T, 1.09 T, and 1.56 T respectively (for **B || c)**. Fig. 2(b) shows the position of the resonance lines versus the tilt angle θ, between the magnetic field and the $Bi_2Te_3$ c-axis (θ = 0 means **B || c**, θ = 90 corresponds to **B || s**). The position of the resonance fields was determined by fitting three Gaussian components to each recorded spectrum, Fig. 3(b).

The position of the three resonance lines shows clear dependence 1/cos(θ) characteristic for the cyclotron resonance of 2D objects, for which the cyclotron frequency depends only on the perpendicular component of the magnetic field. This allows us to assign the spectrum to the cyclotron resonance of the topological surface states.

It is worth noting here that we can clearly exclude the possibility of the resonance being due to the bulk cyclotron resonance, which is expected at 27 mT for **B || c** ($m_c$ = 0.080 $m_0$ assumed) and 20 mT for **B || s** ($m_c$ = 0.058 $m_0$). The 2D quantum well states are either unlikely to be responsible for the observed signal, as they reflect the cyclotron mass of the bulk valence band as well. We can also rule out the bulk plasma-shifted cyclotron resonance due to large difference between the plasma frequency ($10^{12}$ Hz) and experimentally set microwave frequency (9.5 GHz). The coupling of the surface cyclotron resonance to the surface plasma oscillations is also unlikely as it is dependent on the plasmon wave vector and thus on the sample lateral dimension.[26] We did not observe any influence of the sample size on the resonance signal.

The linear dispersion of relativistic fermions results in the Landau level (LL) energy spectrum described by the equation:

$$E_n = \text{sgn}(n) v_F \sqrt{2\hbar |eBn|}, \quad \text{Eq.2.}$$

where n = 0, ±1, ±2, ±3,... is a LL index, $v_F$ denotes the Fermi velocity, $\hbar$ is the Planck constant over 2π, e is electron charge, and B is the magnetic field. In our experiment, the microwave frequency is set constant (f = 9.5 GHz), while the magnetic field is swept in order to tune the transition energy to the energy of microwaves. The condition for the cyclotron resonance is:

$$hf = E_{n+1} - E_n. \quad \text{Eq.3.}$$

The resonance field depends on both the Fermi velocity and the n-index, while the ratio of two resonance fields depends only on n-indexes. This relation allows us to assign L1, L2, and L3 resonance lines to appropriate transitions. The measured ratio of the resonance fields equals to: $B_{L3} : B_{L2}$ = 1.43 ± 0.03 and $B_{L2} : B_{L1}$ = 1.79 ± 0.03 respectively. The calculated ratio for cyclotron resonance transitions between LLs with four lowest indexes is: $B_{|n|=3 \leftrightarrow |n|=4} : B_{|n|=2 \leftrightarrow |n|=3}$ = 1.41 and $B_{|n|=2 \leftrightarrow |n|=3} : B_{|n|=1 \leftrightarrow |n|=2}$ = 1.70. The agreement is very promising taking into account the experimental error in the determination of the resonance field. To sum up, we attribute the L1 line to the cyclotron resonance transition between Landau levels with |n| = 1 and |n| = 2, the L2 line to the transition between |n| = 2 and |n| = 3, and the L3 line to the transition between |n| = 3 and |n| = 4. Once knowing the indexes for particular cyclotron resonance lines we can determine the value of the Fermi velocity which equals to 3260 m/s. A scheme of the energy spectrum of Landau levels in $Bi_2Te_3$ is shown in Fig. 3(a). In the description proposed above we have used a notation "↔" in order to account for both absorption and relaxation processes occurring simultaneously in the resonance cavity.

Based on the observation of the cyclotron resonance with the lowest n-indexes, one can conclude that the Dirac point is located near the Fermi level in sample C. From Shubnikov-de Haas we have already determined that the Fermi level lies 14 meV below the bulk valence band edge. Regarding samples A and B, the Fermi level lies deeper in the bulk valence band (see Tab. 1) which means that it lies also substantially below the Dirac point. This explains the lack of the cyclotron resonance in X-band, because the resonance transitions involving Landau levels with high n-indexes would require magnetic field higher than 40 T, therefore fall out of the range of available magnetic field. Moreover, surface states below the Dirac point seem to be degenerated with bulk valence band states,[5] which may prevent observation of the cyclotron resonance at all.

ARPES determines position of the Dirac point with respect to the bulk band structure, about 130 meV below the $Bi_2Te_3$ valence band edge.[5] The location of the Dirac point close to the Fermi level in sample C requires upward band bending by about 116 meV at the surface. It has been schematically shown in Fig. 3 (c). This is different from Ref. 28 where downward band bending has been observed in $Bi_2Te_3$, $Bi_2(Te_{2.6}Se_{0.4})$, and $Bi_2Se_3$ after short exposition of the sample surface to air. On the other hand in Ref. 8 an upward band bending has been reported for $Bi_2Se_3$. It seems that the problem in $Bi_2Te_3$ and related materials requires further studies. Nevertheless, it becomes clear that not only bulk doping determines occupation of the topological surface states, but surface doping and resulting band bending is not less important.

The most striking conclusion from the cyclotron resonance experiment is the value of the Fermi velocity. ARPES measurements reveal that the surface states of $Bi_2Te_3$ consist of a single, non-degenerate Dirac cone at the Γ point with Fermi velocity of the order of $4 \times 10^5$ m/s,[5] while our cyclotron resonance studies reveal much lower Fermi velocity, equal to 3260 m/s. Still little is known about the nature of topological surface states, thus many explanations of the discrepancies can be considered. We could attribute the difference in the Fermi velocity to imperfectness of the surface morphology. This explanation is however hard to accept, keeping in mind that we deal with topologically protected surface states. On the other hand, similar effects have been recently observed in ultrahigh-mobility $Bi_2Se_3$.[27] Despite a low bulk concentration of the order of $10^{16}$ cm$^{-3}$, no effects of topological surface could be observed in electric transport measurements, possibly due to low mobility of surface electrons. The authors of Ref. 27 conclude that topological protection does not guarantee high surface mobility, and that controlled surface preparation may be necessary in order to obtain desired transport parameters. An alternative explanation is contamination of the surface exposed to ambient conditions. Recent ARPES works on topological insulators from the $Bi_2(Se_{3-x}Te_x)$ family show that the surface states exposed to $N_2$ or air for only 5 min., although demonstrate the robustness of the topological order, they also show the effects of modification of the electronic structure of the surface states. This includes additional doping of electrons into the surface states, modification of the Dirac cone shape (lowering of the Fermi velocity, enhancement of the warping), and finally formation of 2D quantum well states which occurs due to intercalation of gases between quintuple layers.[28] The presented effects of modification of the Dirac cone are not clear enough to explain reduction of the Fermi velocity by two orders of magnitude. However, no data on the long-term exposition to ambient conditions are yet available.

To sum up, we observed the cyclotron resonance in $Bi_2Te_3$ sample exposed to ambient conditions. The nature of the states responsible for the resonance is unambiguously topological with the linear Dirac dispersion. The corresponding Fermi velocity equals to 3260 m/s, which is considerably lower than derived from ARPES[5] for samples cleaved in vacuum. Occupation of topological surface states depends not only on bulk Fermi level but also on surface band bending. The Dirac point is located about 14 meV below the $Bi_2Te_3$ valence band edge in the bulk, which requires upward band bending by about 116 meV at the surface. Observation of the cyclotron resonance in a sample exposed to ambient conditions illustrates exceptional properties of the topologically protected surface states. Surface contamination with atmospheric gases will lower Fermi velocity by two orders of magnitude, it will not however cause the extinction of the surface states. Their nature will remain relativistic.

## ACKNOWLEDGEMENTS

This work has been supported by funds for science, grant number: 2011/03/B/ST3/03362, Poland.

FIGURES:

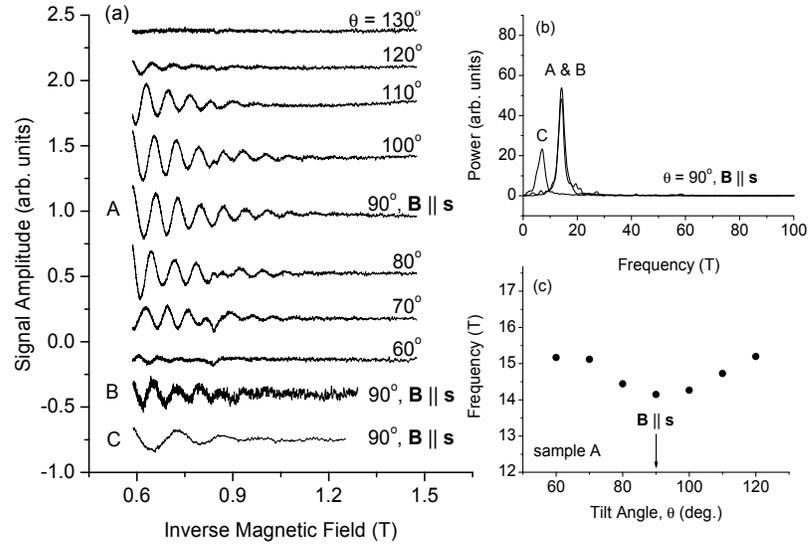

FIG 1. (a) SdH oscillations in p-type $Bi_2Te_3$ recorded in the resonance cavity. For sample A, the spectra were taken versus angle θ between the direction of the magnetic field **B** and $Bi_2Te_3$ **c**-axis. For samples B and C, the spectra at θ = 90° are shown. (b) Fourier transform shows one oscillation frequency equal to 14.2 T, 14.0 T, and 7.0 T for samples A, B, and C, respectively, at θ = 90°. (c) frequency of the SdH oscillations versus tilt angle θ indicates bulk origin.

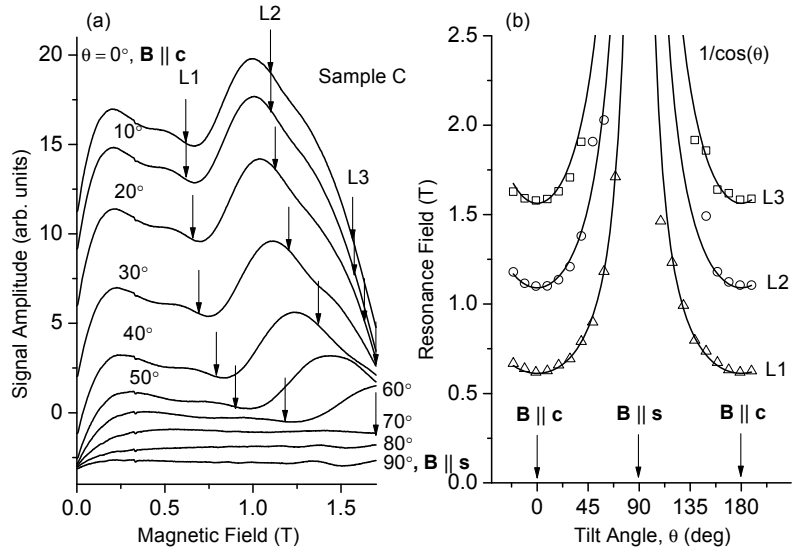

FIG 2. Cyclotron resonance in p-type $Bi_2Te_3$, sample C. (a) spectra recorded versus tilt angle $\theta$ between **B** and **c**. (b) resonance fields of L1, L2, and L3 lines versus the tilt angle $\theta$ – symbols. Solid lines show $1/\cos(\theta)$ dependence. Both the position of the arrows in (a) and the resonance fields in (b) were determined by fitting three Gaussian components to each recorded spectrum, see Fig 3(b).

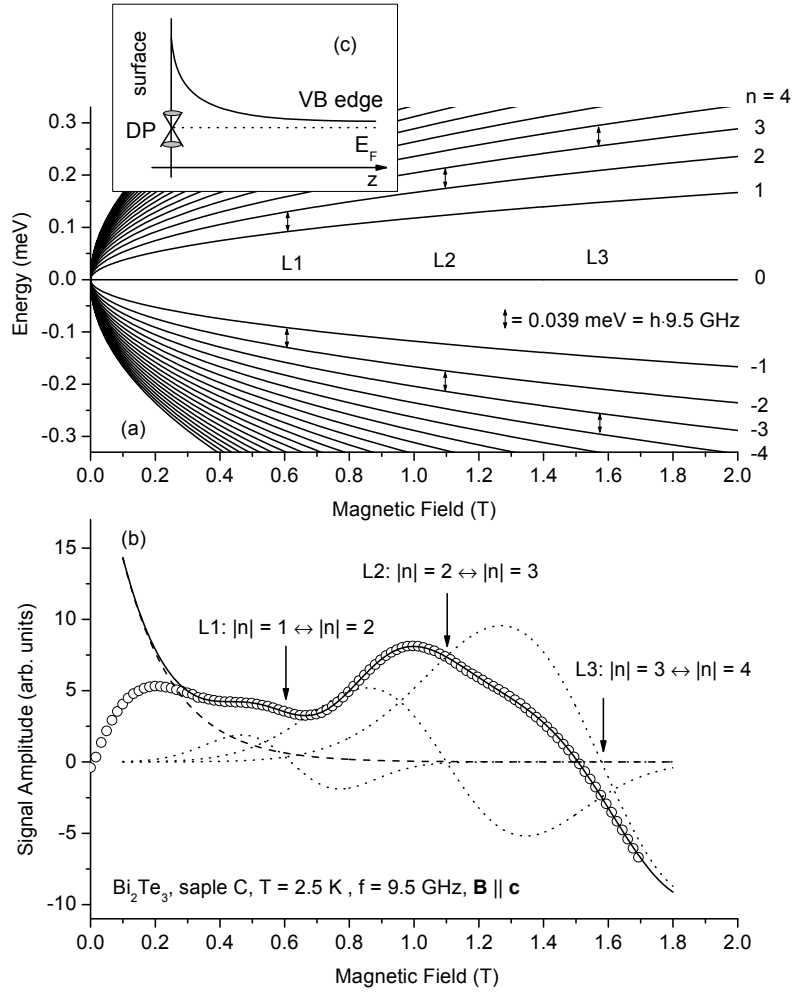

FIG 3. (a) diagram of Landau levels in $Bi_2Te_3$ exposed to ambient conditions. Arrows indicate cyclotron resonance transitions in X-band assigned to L1, L2, and L3 lines respectively. (b) decomposition of the recorded spectrum (open points) into three Gaussian components (dotted lines). Exponential background was assumed (dashed line). Solid line shows the fit. Note that the signal represents first derivative of the microwave power absorption due to the use of the lock-in detection technique. (c) The inset shows schematic representation of band bending at the surface of $Bi_2Te_3$.

TABLE:

TABLE I. Parameters for $Bi_2Te_3$ samples: distance from the crystal seed l, conductivity type, frequency of the SdH oscillations recorded for **B** ∥ **s**, F. $m_{c\ B∥s}$ is the assumed cyclotron mass (after Köhler, Ref. 22) and $E_F$ is the Fermi energy evaluated from Eq. 1 (counted from the bulk valence band edge). Last column says whether the cyclotron resonance in X-band was observed.

| Sample | l (mm) | type | $F_{B∥s}$ (T) | $m_{c\ B∥s}$ ($m_0$) | $E_F$ (meV) | CR |
|---|---|---|---|---|---|---|
| A | 73 | p | 14.2 | 0.075 | 22 | NO |
| B | 80 | p | 14.0 | 0.075 | 22 | NO |
| C | 107 | p | 7.0 | 0.058 | 14 | YES |

[1] C. L. Kane and E. J. Mele, Phys. Rev. Lett. **95**, 146802 (2005).
[2] C. L. Kane and E. J. Mele, Phys. Rev. Lett. **95**, 226801 (2005).
[3] H. Zhang, Ch.-X. Liu, X.-L. Qi, Xi Dai, Z. Fang, and Sh.-Ch. Zhang, Nature Physics 5, 438 (2009).
[4] M. Konig, S. Wiedmann, C. Brune, A. Roth, H. Buhmann, L. W. Molenkamp, X.–L. Qi, and S.-C. Zhang, Science **318**, 766 (2007).
[5] Y. L. Chen, J. G. Analytis, J. H. Chu, Z. K. Liu, S. K. Mo, X. L. Qi, H. J. Zhang, D. H. Lu, X. Dai, Z. Fang, S. C. Zhang, I. R. Fisher, Z. Hussain, and Z.X. Shen, Science **325**, 178 (2009).
[6] D. X. Qu, Y. S. Hor, J. Xiong, R. J. Cava, and N. P. Ong, Science **329**, 821 (2010).
[7] D. Hsieh, Y. Xia, L. Wray, D. Qian, A. Pal, J. H. Dil, J. Osterwalder, F. Meier. G. Bihlmayer, C. L. Kane, Y. S. Hor, R. J. Cava, and M. Z. Hasan, Science **323**, 919 (2009).
[8] J. G. Analytis, J.-H. Chu, Y. Chen, F. Corredor, R. D. McDonald, Z. X. Shen, and I. R. Fisher, Phys. Rev. B **81**, 205407 (2010).
[9] T. Arakane, T. Sato, S. Souma, K. Kosaka, K. Nakayama, M. Komatsu, T. Takahashi, Zhi Ren, Kouji Segawa, and Yoichi Ando, Nature Communications **3**, 636 (2012).
[10] H. Zhang, Ch.-X. Liu, X.-L. Qi, Xi Dai, Z. Fang, and Sh.-Ch. Zhang, Nature Physics 5, 438 (2009).
[11] Z. Alpichshev, J. G. Analytis, J.-H. Chu, I. R. Fisher, Y. L. Chen, Z. X. Shen, A. Fang, and A. Kapitulnik, Phys. Rev. Lett. **104**, 016401 (2010).
[12] Z. Ren, A. A. Taskin, S. Sasaki, K. Segawa, and Y. Ando, Phys. Rev. B **82**, 241306 (2010).
[13] Y. Xia, D. Qian, D. Hsieh, L. Wray, A. Pal, H. Lin, A. Bansil, D. Grauer, Y. S. Hor, R. J. Cava, and M. Z. Hasan, Nat. Phys. **5**, 398 (2009).
[14] M.Z. Hasan and C.L. Kane, Rev. Mod. Phys. **82**, 3045 (2010).
[15] Y. S. Kim, M. Brahlek, N. Bansal, E. Edrey, G. A. Kapilevich, K. Iida, M. Tanimura, Y. Horibe, S.-W. Cheong, and S. Oh, Phys. Rev. B **84**, 073109 (2011).
[16] H. Peng, K. Lai, D. Kong, S. Meister, Y. Chen, X.-Y. Qi, S.-Ch. Zhang, Z.-X. Shen, and Y. Cui, Nature Materials **9**, 225 (2010).
[17] O. E. Ayala-Valenzuela, J. G. Analytis, J.-H. Chu, M. M. Altarawneh, I. R. Fisher, and R. D. McDonald, arXiv: 1004.2311v1 (2010).
[18] A. Hruban, A. Materna, W. Dalecki, G. Strzelecka, M. Piersa, E. Jurkiewicz-Wegner, R. Diduszko, M. Romaniec, and W. Orłowski, Acta Phys. Pol. A **120**, 950 (2010).

[19] A. Wolos, W. Jantsch, K. Dybko, Z. Wilamowski, C. Skierbiszewski, Phys. Rev. B **76**, 045301 (2007).

[20] G. Hendorfer, M. Seto, H. Ruckser, W. Jantsch, M. Helm, G. Brunthaler, W. Jost, H. Obloh, K. Kohler, and D. J. As, Phys. Rev. B **48**, 2328 (1993).

[21] S. J. Youn and A. J. Freeman, Phys. Rev. B **63**, 085112 (2001).

[22] H. Kohler, Phys. Stat. Sol. (b) **74**, 591 (1976).

[23] H. Kohler, Phys. Stat. Sol. (b) **75** 441 (1976).

[24] H. Malissa, Z. Wilamowski, and W. Jantsch, AIP Conf. Proc. 772, pp. 1218-1219, PHYSICS OF SEMICONDUCTORS: 27th International Conference on the Physics of Semiconductors - ICPS-27, 26-30 July 2004, Flagstaff, Arizona (USA).

[25] Z. Wilamowski, N. Sandersfeld, W. Jantsch, D. Tobben, and F. Schaffler, Phys. Rev. Lett. **87**, 026401 (2001).

[26] S. Das Sarma and E. H. Hwang, Phys. Rev. Lett. **102**, 206412 (2009).

[27] N. P. Butch, K. Kirshenbaum, P. Syers, A. B. Sushkov, G. S. Jenkins, H. D. Drew, and J. Paglione, Phys. Rev. B **81**, 241301R (2010).

[28] Ch. Chen, S. He, H. Weng, W. Zhang, L. Zhao, H. Liu, X. Jia, D. Mou, S. Liu, J. He, Y. Peng, Y. Feng, Z. Xie, G. Liu, X. Dong, J. Zhang, X.Wang, Q. Peng, Z. Wang, S. Zhang, F. Yang, Ch. Chen, Z. Xu, Xi, Dai, Z. Fang, and X. J. Zhou, Proc. Nat. Ac. Sci. **109**, 3694 (2012).